\documentclass[prb,superscriptaddress,twocolumn,showpacs]{revtex4}
\usepackage{epsfig}
\usepackage{amsmath}
\usepackage{graphicx}
\begin{document}
\title{Dependence of the vortex structure in quantum dots on the range of the inter-electron interaction}
\author{T. Stopa}
\affiliation{Departement Fysica, Universiteit Antwerpen,
Groenenborgerlaan 171, B-2020 Antwerpen, Belgium}
\affiliation{Faculty of Physics and Applied Computer Science, AGH
University of Science and Technology, al. Mickiewicza 30, 30-059
Krak\'ow, Poland}
\author{B. Szafran}
\affiliation{Departement Fysica, Universiteit Antwerpen,
Groenenborgerlaan 171, B-2020 Antwerpen, Belgium}
\affiliation{Faculty of Physics and Applied Computer Science, AGH
University of Science and Technology, al. Mickiewicza 30, 30-059
Krak\'ow, Poland}
\author{M.B. Tavernier}
\affiliation{Departement Fysica, Universiteit Antwerpen,
Groenenborgerlaan 171, B-2020 Antwerpen, Belgium}
\author{F.M. Peeters}
\affiliation{Departement Fysica, Universiteit Antwerpen,
Groenenborgerlaan 171, B-2020 Antwerpen, Belgium}

\date{\today}

\begin{abstract}
The internal structure of a composite fermion is investigated for a
two dimensional parabolic quantum dot containing three electrons. A
Yukawa screened Coulomb interaction is assumed, which allows us to
discuss the evolution of the electron-vortex correlations from the
Coulomb interaction limit to the contact potential limit. The vortex
structure approaches the Laughlin limit non-monotonically through
the formation of intermediate composite fermions in which a flip of
the spatial orientation of the vortices with respect to the position
of the electrons is observed. Only when we limit ourselves to the
lowest Landau level (LLL) approximation the flip appears through the
formation of an intermediate giant vortex at specific values of the
screening length. Beyond the LLL approximation antivortices appear
in the internal structure of the intermediate composite fermions
which prevent the nucleation of giant vortices. We also studied the
system of five electrons and show that the mechanism of the flip of
the vortex orientation found for three-electron system is reproduced
for higher number of electrons.
\end{abstract}
\pacs{73.21.La,73.43.-f,71.10.Pm} \maketitle
\section{Introduction}
Theoretical interpretation of the fractional quantum Hall
effect,\cite{FQHE} (FQHE) observed at high magnetic field in the
spin-polarized two-dimensional electron gas, is based on the
properties of the Laughlin\cite{Laughlin} wave function. FQHE for
electrons is explained\cite{Jain2} in terms of the integer quantum
Hall effect for composite fermions, i.e. quasi-particles
consisting of electrons with additional even number of bound
vortices (or magnetic field fluxes). The vortices appear as zeros
of the many-electron wave function when its phase changes by
$2\pi$ on a path around this zero. The electron in a composite
fermion feels a reduced effective magnetic field as the bound
vortices partly cancel the usual Aharonov-Bohm phase on a closed
loop around the electron.\cite{wstep} The original problem
considered by Laughlin,\cite{Laughlin} i.e. the diagonalization of
the few-electron eigenequation in the basis of single-electron
wave functions obtained in the symmetric gauge, is formally very
similar to an electron system confined in a parabolic quantum dot.
Only very recently wider attention was paid to the vortices in the
quantum Hall regime of confined
systems\cite{Marten,Saarikoski1,Saarikoski2,Harju,Tober1,Tober2}
and to the composite fermion theory for quantum
dots.\cite{Yan,JQD1,JQD2} In particular, the vortex distribution
for Coulomb interacting electrons confined in quantum dots was
investigated\cite{Marten,Saarikoski1,Saarikoski2} using the exact
diagonalization technique and the reduced wave function imaging.
The structure of vortices as obtained from such exact calculations
differs significantly from the one assumed in the Laughlin wave
function or in the composite fermion approach. It was
found\cite{Marten,Saarikoski1,Saarikoski2} that the vortices are
not localized on the electron as assumed in the Laughlin state but
stay in the neighborhood of electrons to which they are bound. On
the other hand, Laughlin functions are the exact non-degenerate
ground state wave functions for the case of short range
interactions. Analytical proof of their exactness and uniqueness
was provided\cite{Trugman} for potentials developed in series of
$\nabla^{2j}\delta^2({\mathbf r})$. The energy gap allowing the
Laughlin liquid to be incompressible was identified\cite{Haldane}
as due to the short-range component of the Coulomb interaction.

The purpose of the present work is to investigate how the vortex
structure is modified when the inter-electron interaction is taken
from the Coulomb limit to the contact potential limit. We show
that the vortices approach the Laughlin liquid limit in a
non-monotonic fashion. Within the
 lowest Landau level (LLL) approximation for filling factors
$\nu<1/3$ intermediate composite fermion states are found with two
additional vortices localized on the electron. Beyond the LLL
approximation the internal structure of the intermediate composite
fermion turns out to be very complex with possible appearances of
antivortices which can even be localized at the position of the
electron. Within the LLL we found that only in the contact
potential limit more than two extra vortices are localized at the
electron position.

In the present paper we focus our attention on the lowest number of
electrons, i.e. $N=3$, for which a nontrivial\cite{Marten} internal
composite fermion structure can be observed in the reduced wave
function. Next, we verify the conclusions reached for $N=3$ studying
the vortex structure of a five electron system. To study the
dependence of the structure on the range of the inter-electron
interaction we assume that the electrons interact through a Yukawa
potential
\begin{equation}
\label{yukawa} V(r)=\frac{e^2}{4\pi\epsilon_0\epsilon}\frac
{\exp(-r/\alpha)} {r},
\end{equation}
which in the large and small screening length ($\alpha$) limits
yields the Coulomb and the contact potential, respectively. A
potential of the form (\ref{yukawa}) is obtained for an external
Coulomb defect linearly screened by a three dimensional electron
gas.\cite{Ando} In fact the screening of the electron-electron
interaction in electrostatic quantum dots results from charges
induced on the metallic electrodes and is of a more complex
form.\cite{Maksym} The screening of the electron-electron
interaction by the image charges cuts off the long tail of the
Coulomb potential. The contact potential limit corresponds then to
the case of a negligible distance of the quantum dot to the metal
gate in comparison to the dots size.

This paper is organized as follows: Section II presents the theory
behind the results which are given in Section III. Subsection III
(a) contains the results calculated in the LLL approximation and the
influence of the higher LL is described in subsection III (b),
results for five electrons are given in subsection III (c). Summary
and conclusions are provided in Section IV.

\section{Theory}

The effective mass Hamiltonian of our system is
\begin{equation}
\label{ham}
\hat{H}=\sum_{i=1}^N\left(\frac{\left(-i\hbar\nabla_i+e{\mathbf
A}(\mathbf{r}_i)\right)^2}{2m^*}+ V_\text{ext}(r_i)\right)+
\sum_{i<j}^NV(r_{ij}),
\end{equation}
where \mbox{$V_\text{ext}(r)=\frac{1}{2}m^*\omega^2r^2$} is the
parabolic confinement potential, and $\mathbf{A}$ is the vector
potential. We adopt the GaAs effective mass $m^*=0.067m_e$ and
dielectric constant $\epsilon=12.4$. All the calculations were
performed for $\hbar\omega=1$ meV for which the oscillator length
equals $l_0\equiv\sqrt{\hbar/m^*\omega}=33.7$ nm. The
Schr\"odinger equation is solved using the exact diagonalization
(ED) technique\cite{Marten2} with the three-electron Slater
determinants constructed from the single-electron Fock-Darwin
orbitals.\cite{RM} We investigated the ground-state magnetic-field
induced angular momentum and spin transitions of the
three-electron system as function of the screening length in the
presence of a perpendicular magnetic field
[$(0,0,B)=\nabla\times\mathbf{A}$].
 For $\alpha\rightarrow\infty$ we
exactly reproduce the results of Ref.[\cite{Mikha}] (our
parameters correspond to the interaction constant $\lambda\equiv
l_0/a_B=3.44$, with $a_B$ the donor Bohr radius). For finite
values of $\alpha$ no interesting results are obtained: decreasing
the screening length has the trivial effect of decreasing the
strength of the interaction ($\lambda$), the ground-state
spin-orbital symmetry sequence remains unchanged, only the
critical magnetic fields for the transitions between subsequent
angular momentum states are shifted to higher values.

We consider only the spin-polarized states of the magic angular
momentum sequence\cite{RM} [total angular momentum $L\hbar$ being
a multiple of $3\hbar$], which become ground states at high
magnetic fields, after the maximum density droplet decays. The
results presented below were obtained mostly within the LLL (more
precisely in the lowest Fock-Darwin band\cite{RM} of zero radial
quantum number and nonnegative angular momentum) to keep a direct
correspondence to the Laughlin wave function. In the discussion of
the vortices we do not apply any magnetic field to the system
without loss of generality for the wave function, since for a
harmonic confinement potential the magnetic field simply rescales
the electron coordinates of the wave function for a given
$L$:\cite{Marten2}
\begin{equation}
\Psi_{B\neq0}(\mathbf{r}_1,\mathbf{r}_2,\mathbf{r}_3)=
\Psi_{B=0}(\gamma\mathbf{r}_1,\gamma\mathbf{r}_2,\gamma\mathbf{r}_3),
\end{equation}
with the scaling factor $\gamma=(1+(\omega_c/2\omega)^2)^{1/4}$,
where $\omega_c=eB/m^*$ stands for the cyclotron frequency.
 Note, that property
(3) implies that if, as generally accepted, the ground states at
high magnetic fields are well approximated by the LLL, the
approximation is not any worse at $B=0$, where the high $L$ states
correspond to high excitations. Moreover, the eigenstates of the
Hamiltonian written in the basis of Slater determinants built of
LLL wave functions can be exactly identified with the eigenstates
of the electron-electron interaction matrix operator. They are
therefore the same for any constant $\lambda$ multiplying the
interaction potential [Eq. (1)], even if for large $\lambda$ the
LLL approximation can be arbitrarily bad.\cite{mikha2} In the
calculations we consider screening lengths $\alpha\geq0.1$ nm. The
delta-like interaction potential obtained for $\alpha\rightarrow
0$ does not influence the energies or wave functions for a
spin-polarized system because of the Pauli exclusion principle.
Consequently, for $\alpha=0$ one obtains a multifold degenerate
non-interacting ground-state. Since in the diagonalization these
states (with very different vortex structure each) mix
stochastically one cannot carry on the discussion of vortices for
a screening length equal to zero. As a matter of fact, there is
actually no need to take $\alpha$ strictly zero, since then the
Laughlin function, as well as any other wave function constructed
within the LLL, obviously corresponds to the degenerate ground
state.

\begin{figure*}[htbp]
\hbox{\hbox{\epsfysize=46mm
                \epsfbox[26 88 560 690] {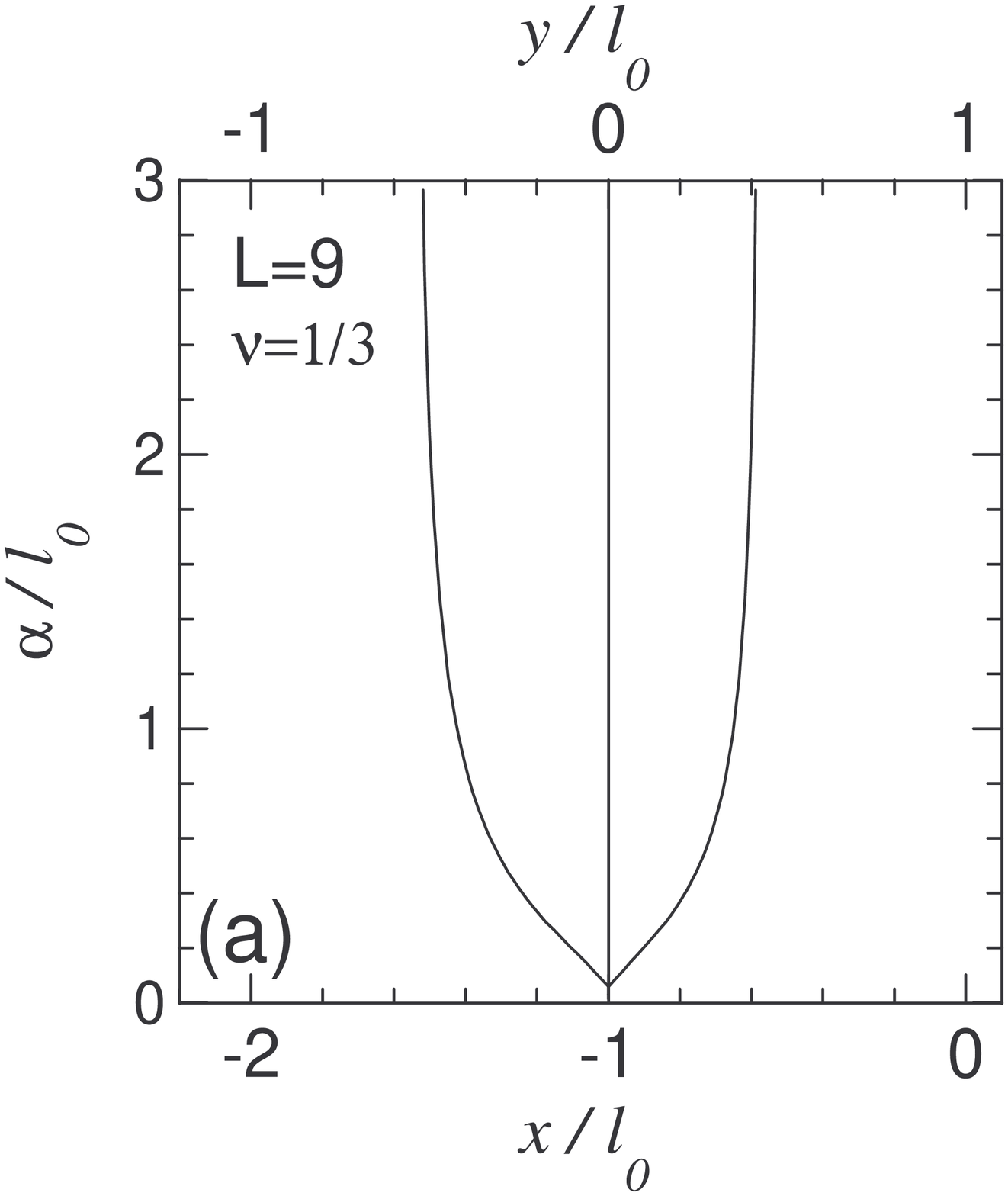}\hfill}
                \hbox{\epsfysize=46mm
                \epsfbox[152 88 560 690] {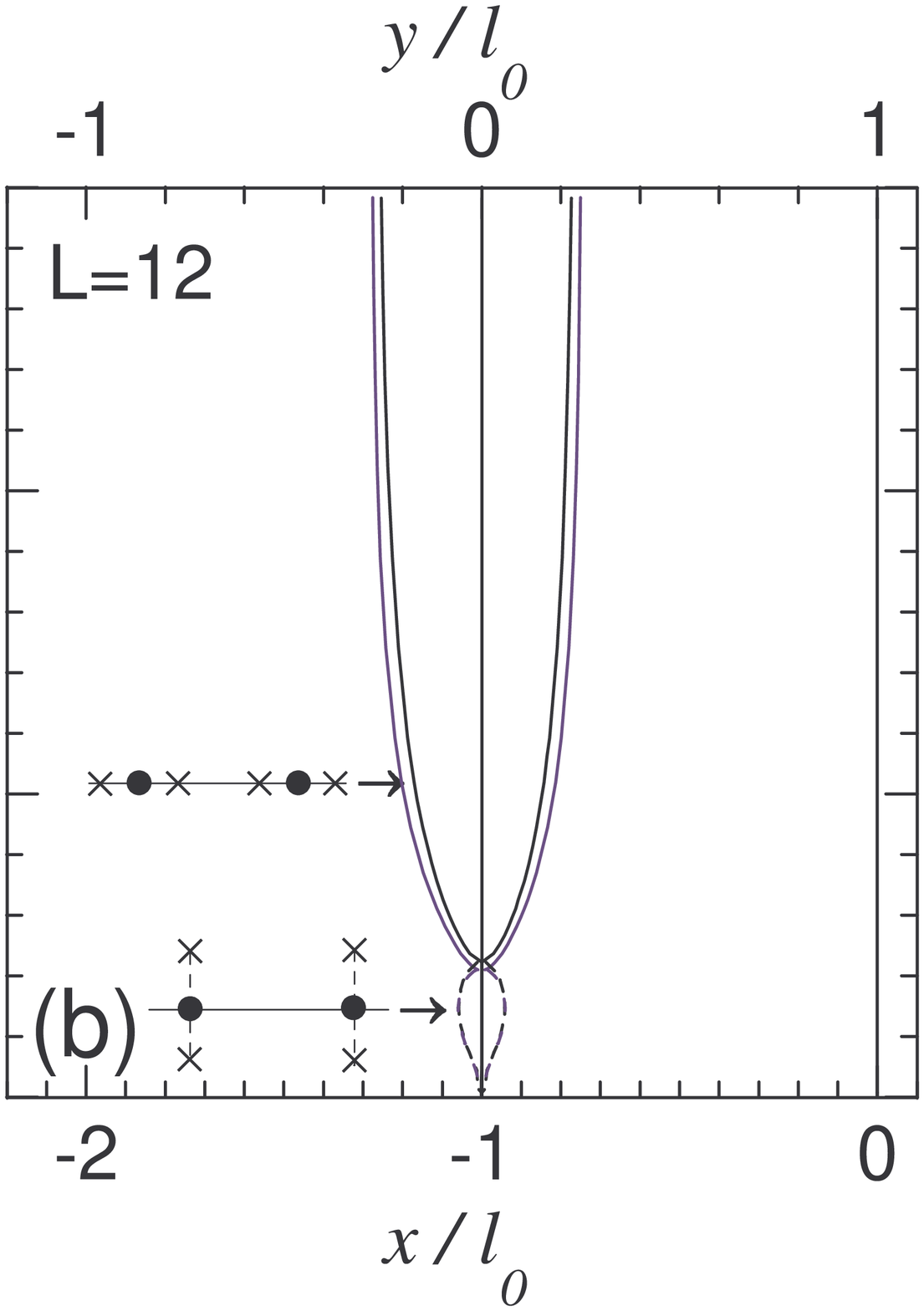}}\hfill
                \hbox{\epsfysize=46mm                 \epsfbox[142 88 560 690] {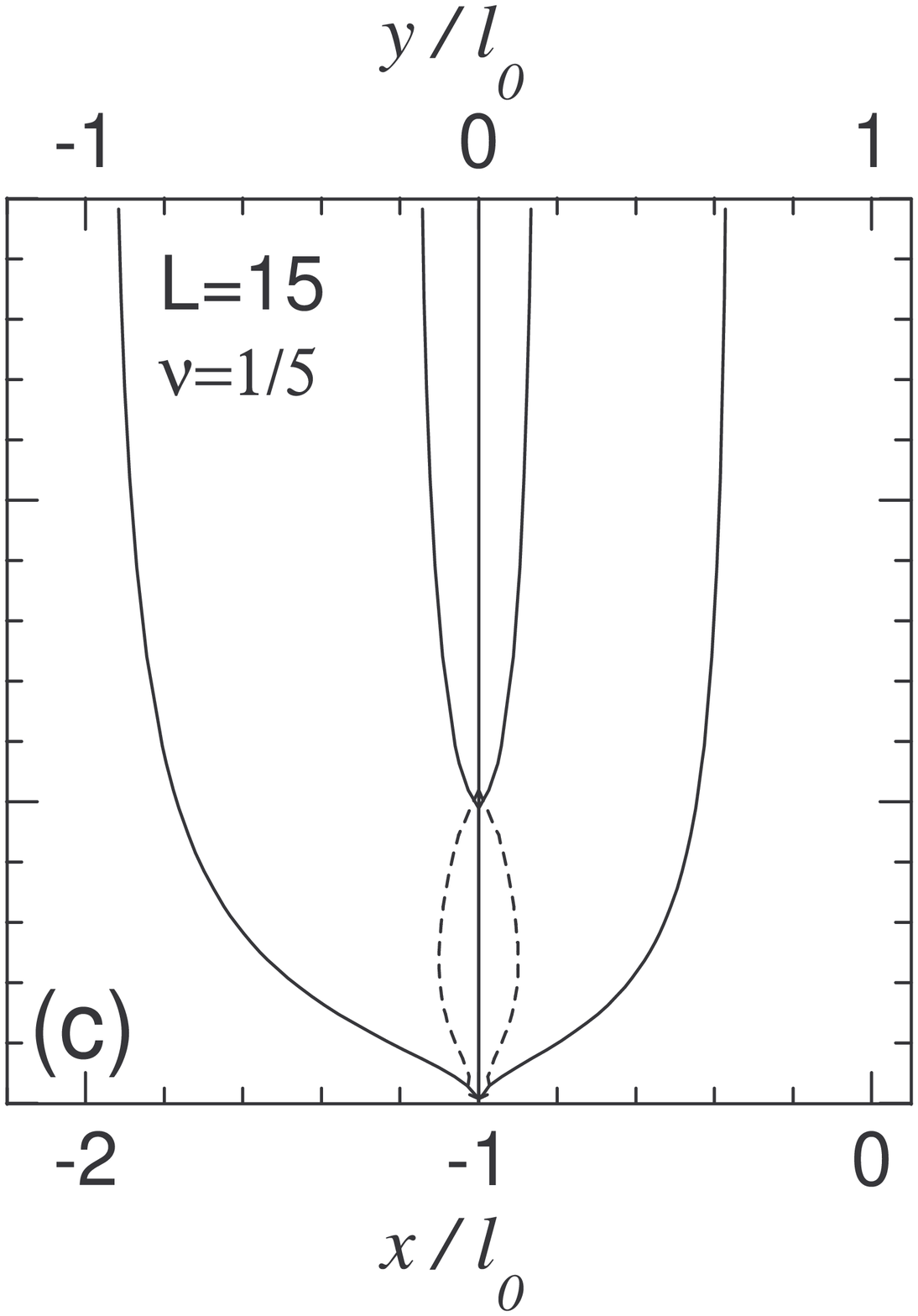}}\hfill
                \hbox{\epsfysize=46mm                 \epsfbox[142 88 560 690] {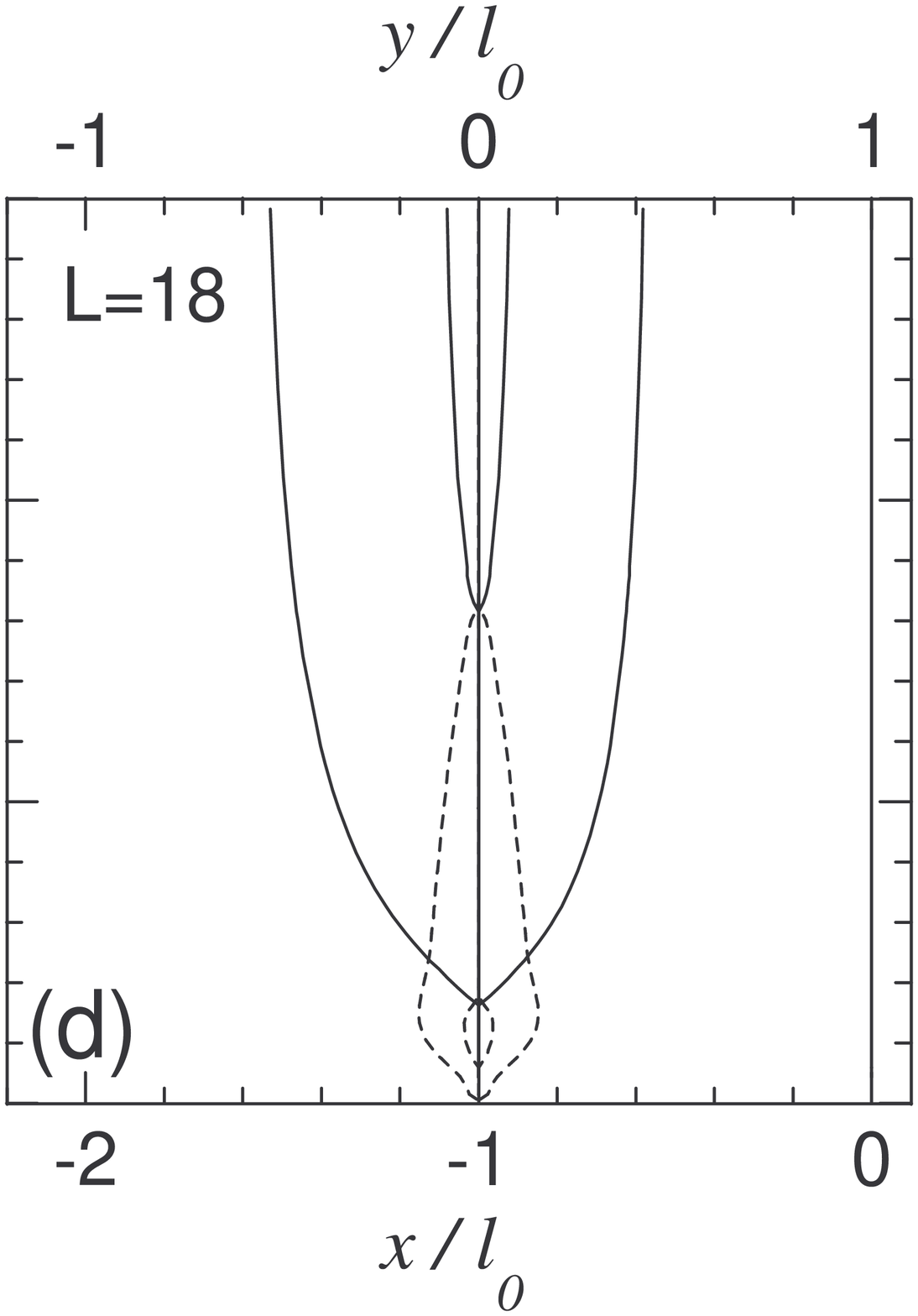}}\hfill
               \hbox{\epsfysize=46mm                 \epsfbox[142 88 560 690] {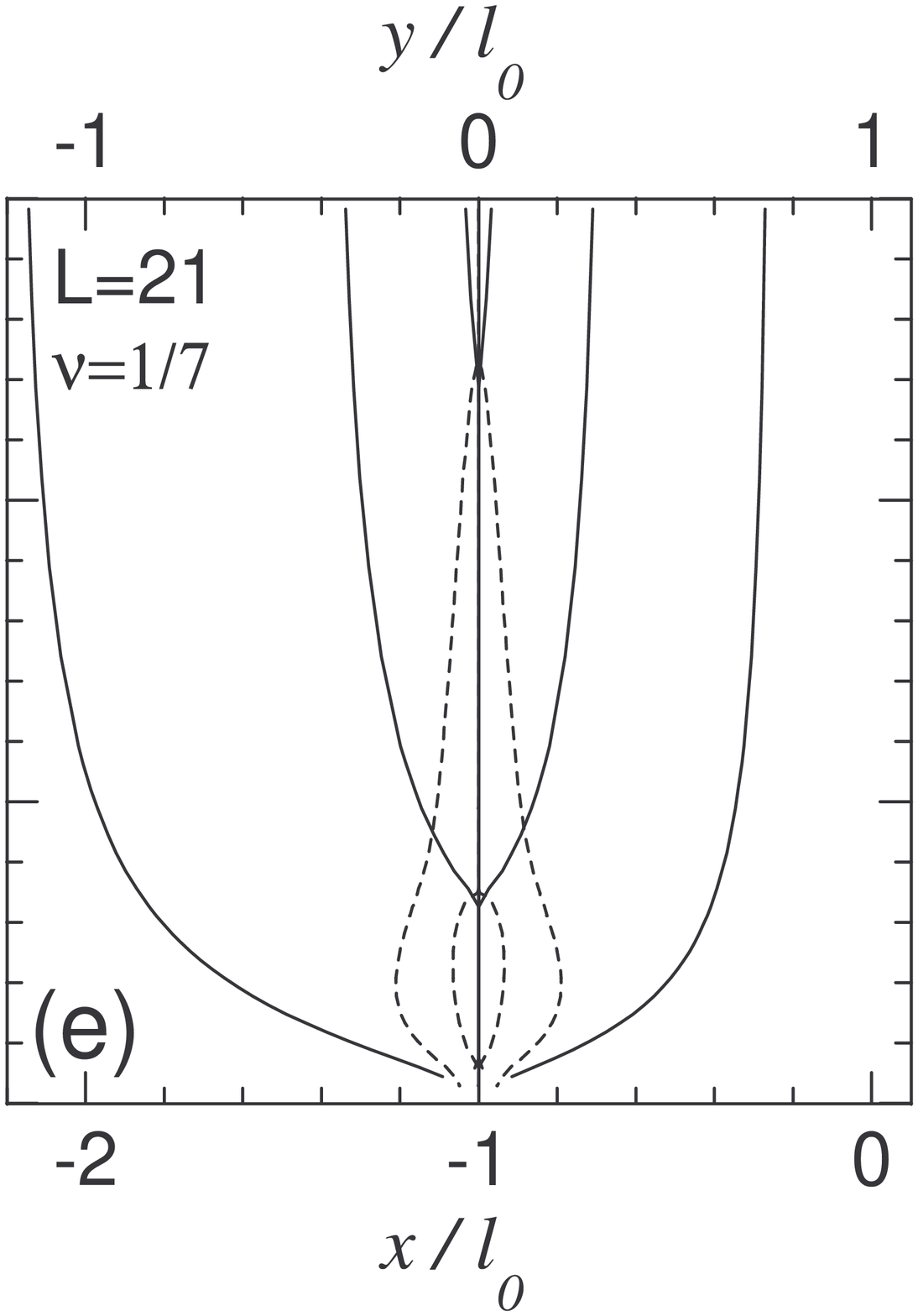}}\hfill
                }

\caption{(color online) Positions of vortices of the conditional
wave function calculated for the two electrons fixed at points $(\pm
l_0,0)$ as function of the screening length $\alpha$. Solid lines
correspond to lower horizontal axis and show the positions of
vortices on the $x$ axis ($y=0$). Dashed lines are plotted with
respect to the upper horizontal axis and show the $y$ coordinate of
vortices localized on the $x=-l_0$ line. All the results were
obtained in the lowest Landau level (LLL) with the exception of the
blue curves plotted in (b) calculated beyond the LLL approximation
with a fully convergent basis set. At the left side of (b) we show
the electron-vortex orientation before ($\alpha>0.44l_0$) and after
the formation of the giant vortex ($\alpha<0.44l_0$). The electron
positions are marked with dots, and the vortices by crosses.}
\label{rys1}
\end{figure*}

\begin{figure}[htbp] \hbox{\epsfxsize=54mm
                \epsfbox[10 230 365 830] {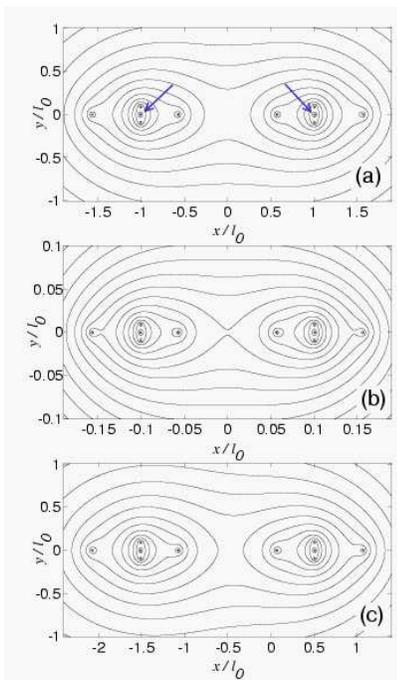}\hfill}
\caption{(color online) Contour plots of the logarithm of the
absolute value of the reduced wave function calculated for angular
momentum $L=15$ and the screening length $\alpha=l_0/2$ in the LLL
approximation for the pinned electron positions $(\pm l_0,0)$ (a),
$(\pm l_0/10,0)$ (b) and $(\pm l_0+l_0/2,0)$ (c). Electron
positions are marked by blue arrows in (a).} \label{scalability}
\end{figure}

A general form of the three-electron wave function in the LLL
approximation is:\cite{uwaga}
\begin{equation}
\label{general} \Psi(z_1,z_2,z_3)=\sum_{j}\eta_j {\text{\em
A}}z_{1}^{j_{1}}z_{2}^{j_{2}}z_{3}^{j_{3}}
\exp\left(-\frac{1}{2}\sum_{k=1}^3\frac{|z_k|^2}{l_0^2}\right)
\end{equation}
where {\em A} stands for the antisymmetrizer, $z_k\equiv x_k+iy_k$
denotes the complex, two-dimensional position of the $k$-th
particle,
 $\eta_j$ are the linear variational parameters,
${j_{1}}$, $j_{2}$, $j_{3}$ are nonnegative integers, of which not
a pair is identical, and $j_1+j_2+j_3=L$.  The Laughlin wave
function\cite{Laughlin} for the angular momentum $L=3m$ (for odd
$m$) is a product of the Jastrow factor and a Gaussian
\begin{equation}
\label{laughlin}
\Phi_{1/m}(z_1,z_2,z_3)=\prod_{k<l}(z_{k}-z_{l})^m\exp\left(-\frac{1}{2}\sum_{n=1}^3\frac{|z_n|^2}{l_0^2}\right),
\end{equation}
which is a special form of the general formula (\ref{general}). In
the Laughlin function the filling factor $\nu=1/m$ is directly
related to the number of zeros $m$ localized on each electron as
well as to the angular momentum \mbox{$\nu=N(N-1)/(2L)$} (for
$N=3$ electrons one has $\nu=3/L$).
 Note that not all the states of the magic
angular momentum sequence can be represented by the Laughlin
function, only those of odd $L$ can.

We investigate the zeros of the reduced wave
function,\cite{Marten,Saarikoski1,Saarikoski2} constructed by
fixing coordinates of two electrons $z_1$ and $z_2$
\begin{equation}
\psi_{z_1,z_2}(z)=\Psi(z_1,z_2,z),
\end{equation}
where $z$ is the test electron coordinate. The reduced Laughlin
wave function is a complex polynomial of the test electron
position ($z$) of degree $2m=\frac{2}{3}L$, multiplied by a
Gaussian.  On the other hand, for a general LLL state
(\ref{general}) the reduced wave function is a complex polynomial
of degree $L-1$, resulting in more zeroes than occurring in the
Laughlin wave function. The additional zeroes, commonly attributed
to vortices bound to the test electron, are not localized close to
the pinned electron positions. Since one extra zero has to be
attributed to the test electron itself, one obtains the total
number of $L$ vortices [for a general number of $N$ electrons the
number of vortices equals $N/\nu=2L/(N-1)$]. When higher Landau
levels are included the reduced wave function depends also on the
complex conjugate of the particle positions and larger exponent
values in the polynomial appear, which increases the number of
zeros and allows antivortices to appear.\cite{wstep,Marten}

\section{Results}
\subsection{Lowest Landau level approximation}

 Fig. 1 shows the position of zeros of the LLL
reduced wave functions for two of the electrons pinned in the
locations $(\pm l_0, 0)$ for states with $L=9, 12, 15, 18$ and $21$.
In the presented range of $x$ we focus only on the zeros located
near the pinned electrons. In the Coulomb limit all the vortices are
placed on the $x$ axis.\cite{Marten} For $L=9$ [Fig. 1(a)], as the
screening length decreases, the two vortices bound to the electron
approach its position, staying always on the $x$ axis. For
$\alpha=2$ nm ($\alpha=0.06 l_0$) the bound vortices are localized
exactly at the electron position forming a giant vortex,
characteristic for the Laughlin wave function. For $L=12$ [see the
black lines in Fig. 1(b)] the giant vortex on the electron position
is formed earlier, i.e. for $\alpha=0.44l_0$. However, for smaller
screening lengths the two extra vortices leave the electron position
(and the $x$ axis) passing to the $x=-l_0$ line [see the inset in
Fig. 1 (b)]. For still smaller $\alpha$ the vortices return to the
electron position. A similar behavior is found for larger $L$. For
states with $L>12$, there are more than $2$ extra vortices bound to
each electron and pairs of them collapse on the electron positions
for specific $L$-dependent screening length values. Decreasing
$\alpha$ beyond this value flips them to the $x=\pm l_0$ line
approaching again the electron positions in the $\alpha=0$ limit.
 Note, that the formation of
the intermediate giant vortices is observed also for non-Laughlin
states (even $L$) and that all these intermediate giant vortices
have winding number three. For non-Laughlin states the number of
zeros of the reduced wave function ($L-1$) is odd, therefore a
single vortex resides in the $(0,0)$ position in order not to break
the symmetry. The position of this vortex for $L=12$ and $L=18$ is
marked by the vertical line just to the left of the tick marks on
the right hand side. We see that for even $L$, i.e. the non-Laughlin
states, the number of vortices bound to the electrons is the same as
in the closest Laughlin state with lower angular momentum ($L-3$).

The presented results are quite general in the sense that for
$N=3$ the vortex structure does not depend on the specific choice
of the positions of the two fixed electrons in Eq. (6) but scales
linearly as a function of the distance between the fixed electron
positions, whether they are, or not, placed symmetrically with
respect to the origin. This is demonstrated in Fig. 2, which shows
a contour plot of the logarithm of the absolute value of the
reduced wave function calculated in the LLL approximation for
$L=15$ and the screening length $\alpha=l_0/2$. In Fig. 2(a) the
two electrons are fixed at $(\pm l_0,0)$ like in Fig. 1 (c). For
$\alpha=l_0/2$ two of the bound vortices are localized
perpendicular to the line between the electrons [cf. Fig. 1(c)]
whose positions  are marked with the blue arrows. In Fig. 2(b) the
fixed electron coordinates were scaled down 10 times with respect
to Fig. 2(a), and in Fig. 2(c) the electrons were shifted to the
left by $l_0/2$. The scalability of the vortex structure is
evident from the form of the LLL wave function (4). If one
decreases all the distances $\gamma$ times, one can take
$\gamma^L$ before the sum (since $j_{1}+j_{2}+j_{3}=L$) from the
polynomial part, i.e. the one responsible for the appearance of
the vortices. The invariance of the vortex structure with respect
to the shift of the fixed electron positions is not evident from
Eq. (4). In fact, the vortex structure of each of the Slater
determinants is {\em not} invariant with respect to the shifts,
and the invariance is only obtained in the entire basis containing
all the LLL determinants. Since we are dealing with the harmonic
oscillator potential the exact wave function is separable into a
product of the center of mass and relative motion wave functions
$\Psi=F_{CM}(z_{cm})G_{rel}(z_1-z_2,z_1-z_3,z_2-z_3)$. The
vortices are entirely due to the relative part. From the separable
form it is clear that the vortices shift with the fixed electron
positions. It is quite remarkable that this feature of the exact
solution is reproduced in the LLL approximation. For a general $N$
the vortex structure remains invariant with respect to the size,
position and orientation of the polygon formed by the $N-1$ fixed
electrons as long as its shape is preserved. One cannot change the
shape of the line segment linking the two fixed electrons for
$N=3$. But for $N=4$ different vortex structures are obtained when
the shape of the triangle formed by the fixed electrons is
varied.\cite{Marten} Fig. 2 shows also that only the vortex
structure and {\em not} the reduced wave function scales with the
positions of the fixed electrons. The nonscalability of the wave
function results from the center of mass component of the wave
function [or the Gaussian in Eq. (4)]

The dashed lines in the upper part of Fig. 3 show the positions of
the two remaining vortices, which did not fit into Fig. 1(a) for
$L=9$. These vortices are not bound to the electrons whose
positions are pinned but belong to the test electron and disappear
to infinity as the screening constant is decreased to zero. This
behavior is expected since the number of vortices in the Coulomb
problem is larger than in the limit of the Laughlin liquid (see
the end of Section II). The red full curves close to the lower
horizontal axis show the modulus of the corresponding reduced
Laughlin wave function for $y=0$ with the electron positions fixed
at $(\pm l_0,0)$. We see that the disappearing test-electron
vortices are always localized beyond the region of the reduced
Laughlin wave function localization. This is not always the case.
Black solid lines in Fig. 3 show the position of vortices for the
two electrons fixed at $(-l_0,0)$ and $(-l_0/2,0)$. The outermost
vortices are localized more closely to the electrons. We see that
in this case, for decreasing screening lengths the vortices pass
through the region in which the reduced Laughlin function (plotted
in blue in Fig. 3) takes large values.

In order to get an idea how well the electron-vortex correlations
are described in the Laughlin wave function we project the reduced
optimal wave function obtained within the LLL approximation to the
one corresponding to the Laughlin many-particle wave function
\begin{equation}
S_{z_1,z_2}=\frac{<\psi_{z_1,z_2}|\psi^L_{z_1,z_2}>}{\sqrt{<\psi_{z_1,z_2}|\psi_{z_1,z_2}><\psi^L_{z_1,z_2}|\psi^L_{z_1,z_2}>}},
\end{equation}
in which the positions of vortices as well as of the pinned
electrons can be seen ($\psi^L$ denotes the reduced Laughlin wave
function). The overlaps calculated between ED wave functions of
states with odd angular momentum and corresponding Laughlin wave
functions are shown in Fig. \ref{over}. In Fig. \ref{over} (a) the
two pinned electrons were placed at $(\pm l_0,0)$. Vortex
positions for this case are shown in Figs. \ref{rys1}(a), 1(c),
and 1(e). The overlap values increase monotonically for all the
three Laughlin states with decreasing $\alpha$, in the
$\alpha\rightarrow 0$ limit they all achieve unity. Note also that
the higher the angular momentum the smaller the overlap. In Fig.
\ref{rys1} we observe that for these three states the distance
between the outermost bound vortex and the electron increases with
$L$ which is the reason why for larger $L$ the overlaps are
smaller. Moreover, there are regions for $L=15$ and $L=21$ where
vortices increase their distance from the electron with decreasing
$\alpha$, which is not reflected in the overlap plot, which
apparently is more strongly determined by the decreasing distance
of the electron from the outermost vortex.

\begin{figure}
\includegraphics[angle=-90, scale=0.35]{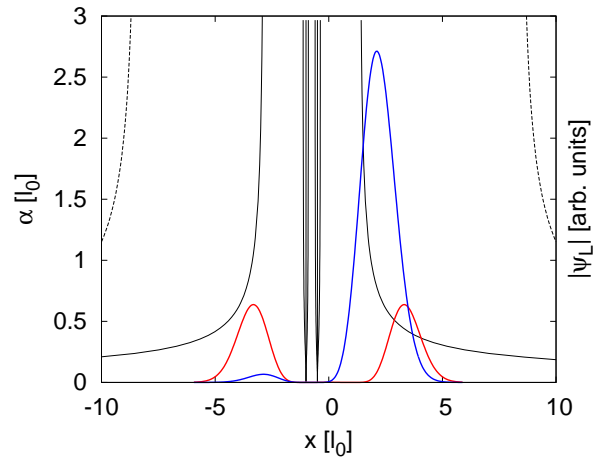}
\caption{(color online) Vortices and the conditional probability
for $L=9$ when the two electrons are pinned at $(-l_0,0)$ and
$(-l_0/2,0)$ (black solid lines). The two outermost vortices for
electrons in pinned at $(\pm l_0,0)$ are shown with dashed curves.
The reduced Laughlin wave functions along the $y=0$ axis are also
shown by red and blue lines, for the symmetrically and
nonsymmetrically pinned electrons, respectively. } \label{l9}
\end{figure}

\begin{figure}[htbp]
\hbox{\epsfxsize=79mm
                \epsfbox[7 75 633  355] {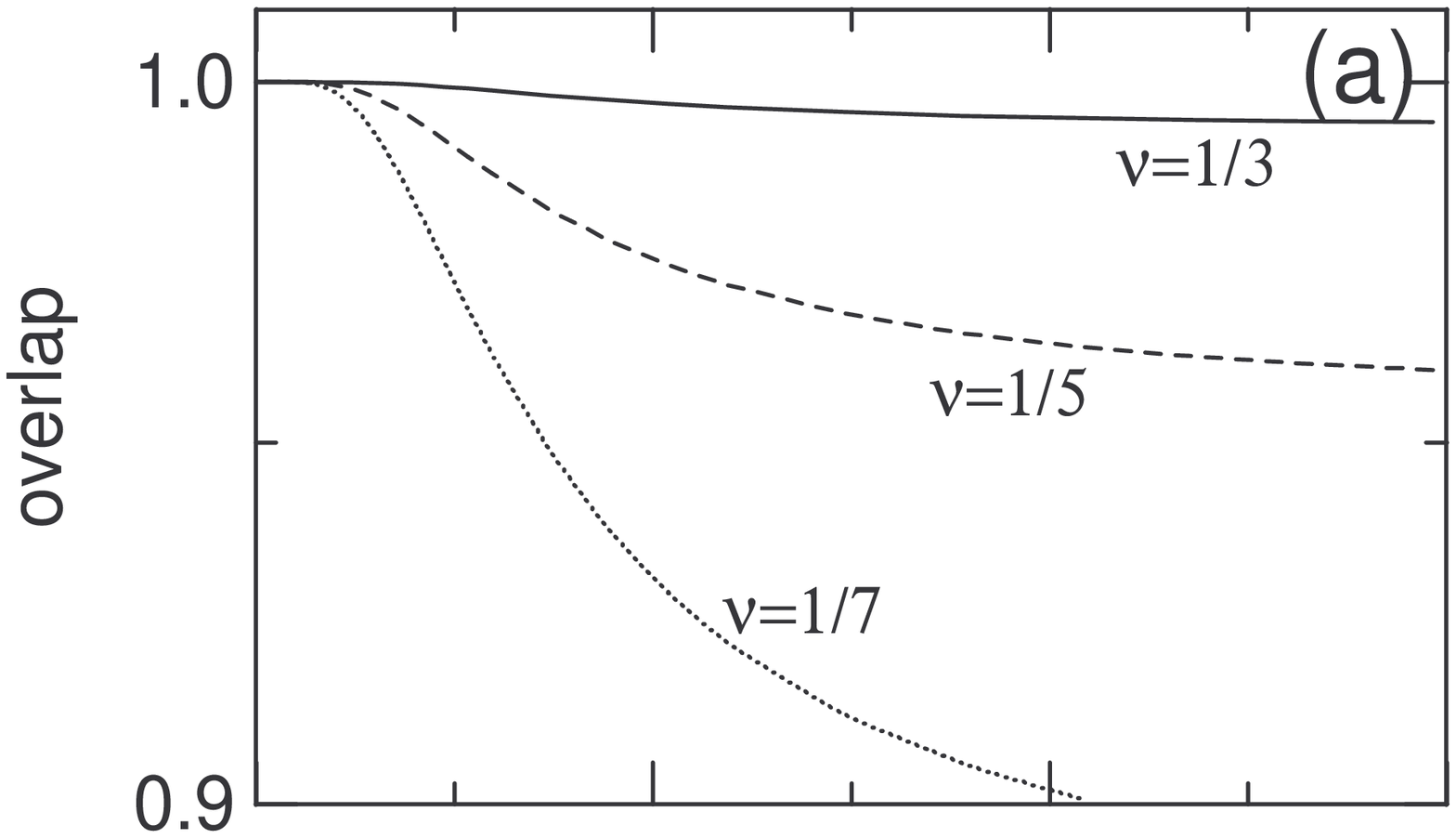}\hfill}

                \hbox{\epsfxsize=79mm
                \epsfbox[2 75 629 421] {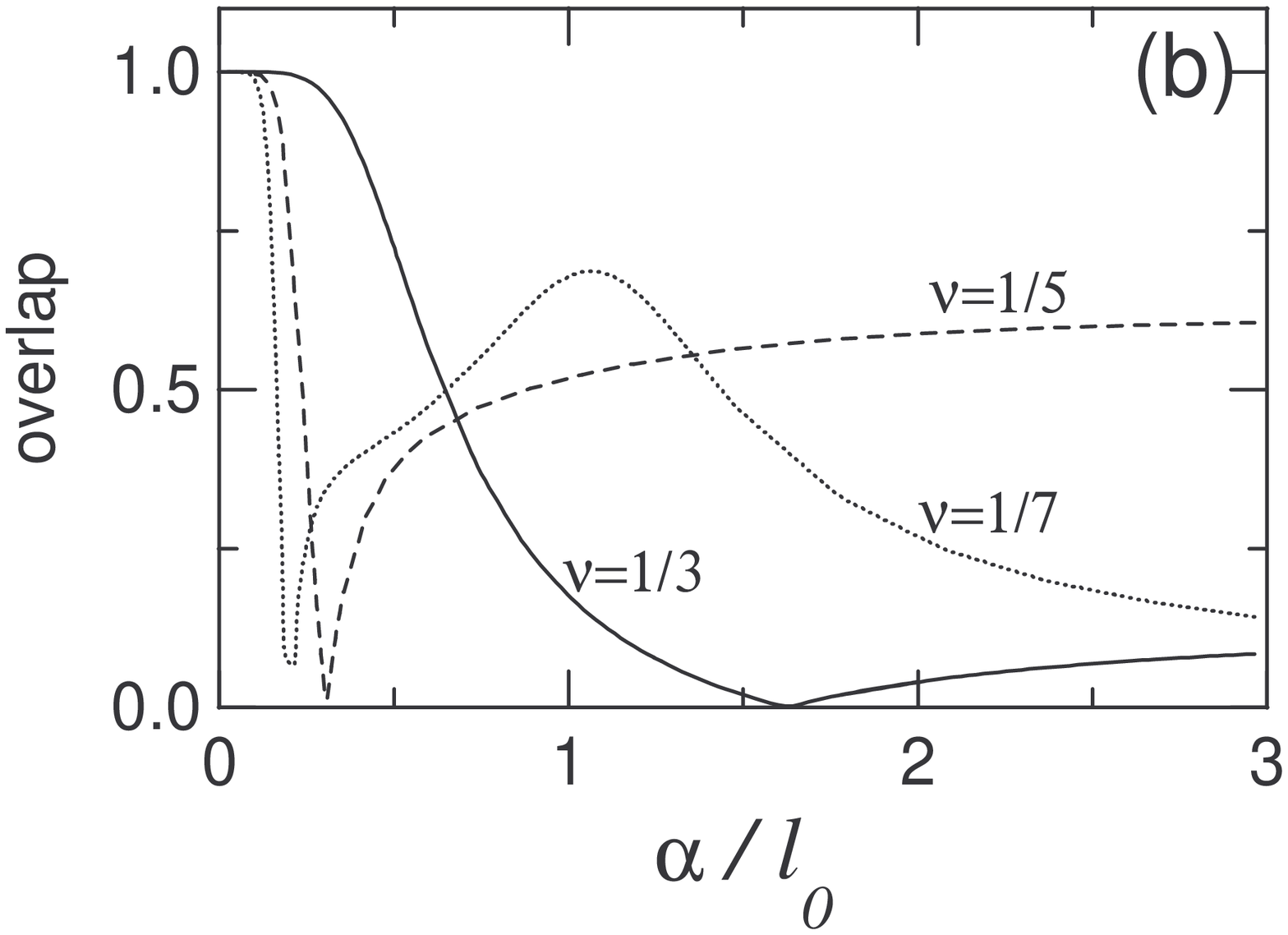}}

\caption{Overlap integral of the conditional wave functions
calculated for the lowest-energy state diagonalizing the
Hamiltonian in the LLL subspace and the state described by the
Laughlin wave function (\ref{laughlin}). In (a) two of the
electrons are fixed at positions $(\pm l_0,0)$, and in (b) at
 $(-l_0,0)$ and $(-l_0/2,0)$.} \label{over}
\end{figure}

More interesting behavior is observed when the electrons are
pinned closer to each other.  We placed them in $(-l_0,0)$ and
$(-l_0/2,0)$ and the resulting overlaps are shown in Fig.
\ref{over} (b). For all three states there is a more sharp minimum
as function of $\alpha$. This is due to the external vortex
passing through the region in which the reduced wave function is
large as discussed in the context of Fig. \ref{l9}. The minimal
value of the overlap, which is almost zero, occurs exactly when
the vortex position coincides with the Laughlin wave function
maximum. The distinctly different dependence of the overlaps for
the fixed-electron positions is due to the fact that the reduced
wave function is not scalable with the interelectron distances.

\begin{figure}[htbp]
\hbox{\epsfxsize=74mm
                \epsfbox[53 324 519 793] {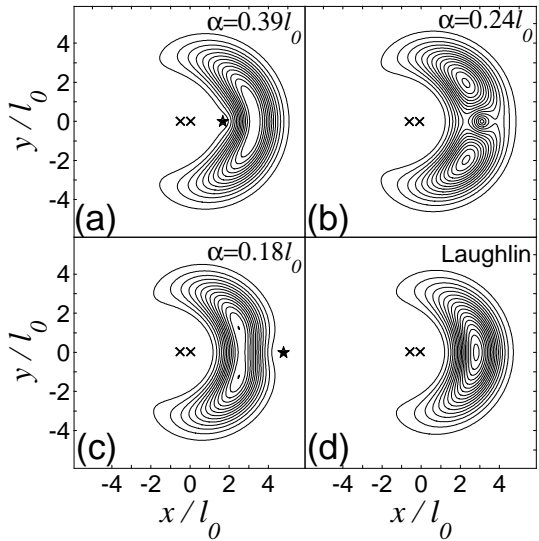}\hfill}
\caption{Contour plots of the absolute value of the reduced wave
function for the state with $L=15$ and the two electrons fixed at
$(-l_0, 0)$ and $(-l_0/2, 0)$ (indicated by crosses). (a)-(c) show
the ED wave function as obtained within the LLL approximation for
different values of $\alpha$, (d) shows the Laughlin wave function
corresponding to this state. One of the vortices bound to the test
electron which crosses through the wave function's maximum is
indicated by a star in (a) and (c), while in (b) it creates a
distinct minimum at $x=3l_0$.} \label{wave1}
\end{figure}

\begin{figure}[htbp]
\hbox{\epsfxsize=80mm
                \epsfbox[60 356 486 815] {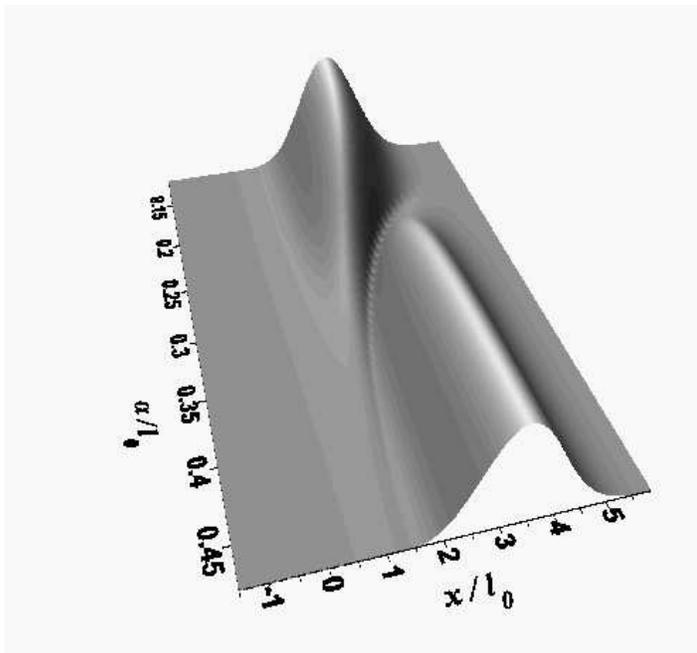}\hfill}
\caption{Evolution of the absolute value of the reduced wave
function at $y=0$ as function of $\alpha$ for the state with
$L=15$ and two electrons pinned in $(-l_0,0)$ and $(-l_0/2,0)$. }
\label{wave2}
\end{figure}

The displacement of the vortex towards infinity for
$\alpha\rightarrow0$ and its effect on the reduced LLL wave
function is illustrated in the contour plots of Fig. \ref{wave1}
for $L=15$ with the fixed electron positions $(-l_0,0)$ and
$(-l_0/2,0)$. The position of the vortex is marked by a $\star$ in
Figs. 5(a) and 5(c). In Fig. 5(b), $\alpha$ corresponds to the
overlap minimum [cf. Fig. 4(b)] the vortex is visible near
$x=3l_0$ where it digs a hole in the wave function. Moreover, when
the vortex is at the position of the Laughlin wave function
maximum, it splits the LLL wave function into two almost equal
parts with opposite signs (see Fig. \ref{wave2}). This makes the
Laughlin function almost orthogonal to the ED wave function. When
the vortex of the test electron passes beyond the maximum of the
wave functions, the overlap starts to increase reaching unity for
all the states, but this occurs earlier for smaller values of $L$.

\begin{figure}[htbp]
\hbox{\epsfxsize=59mm
                \epsfbox[16 214 545 800] {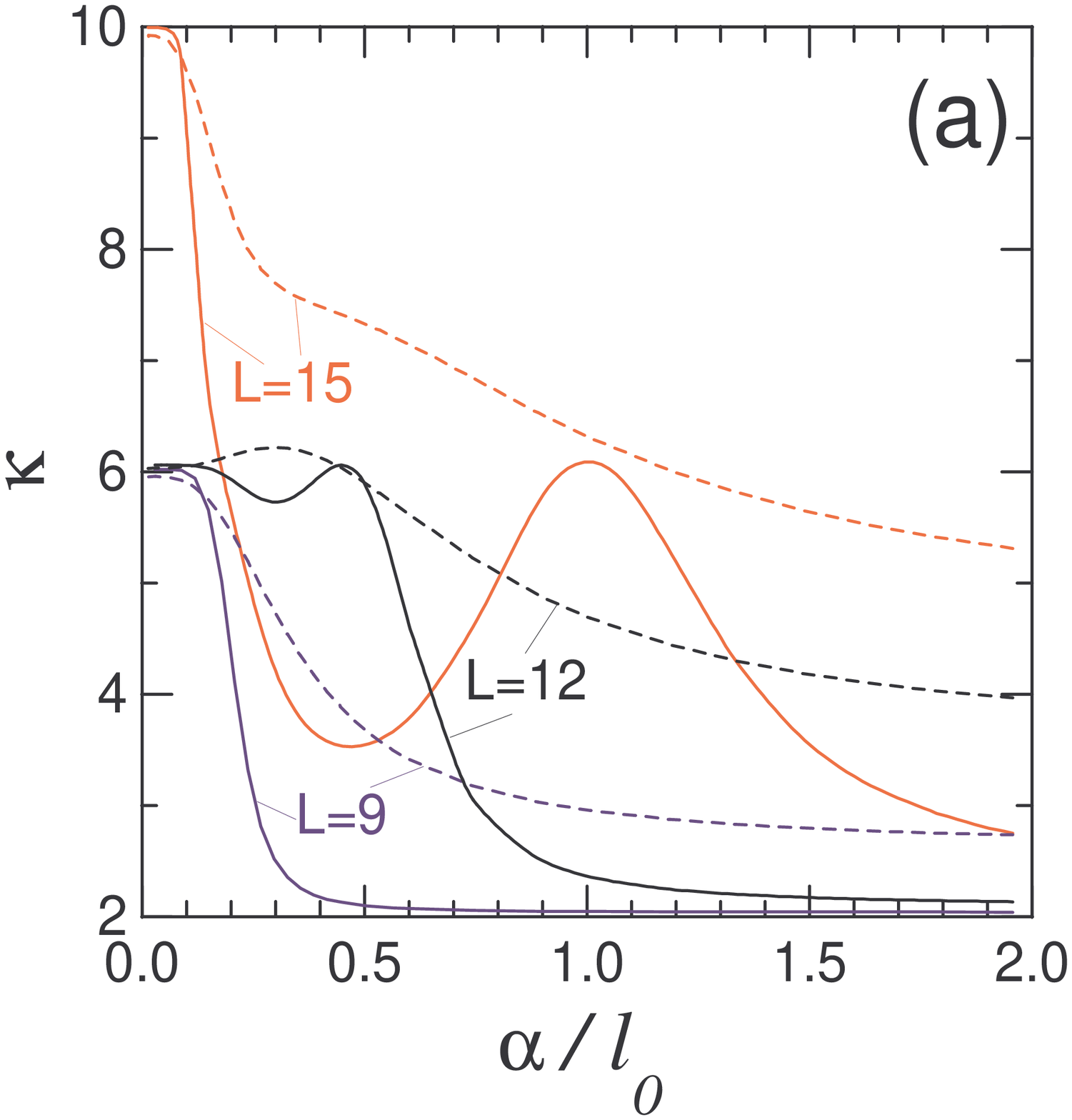}\hfill}
                \hbox{\epsfxsize=59mm
                \epsfbox[16 214 545 800] {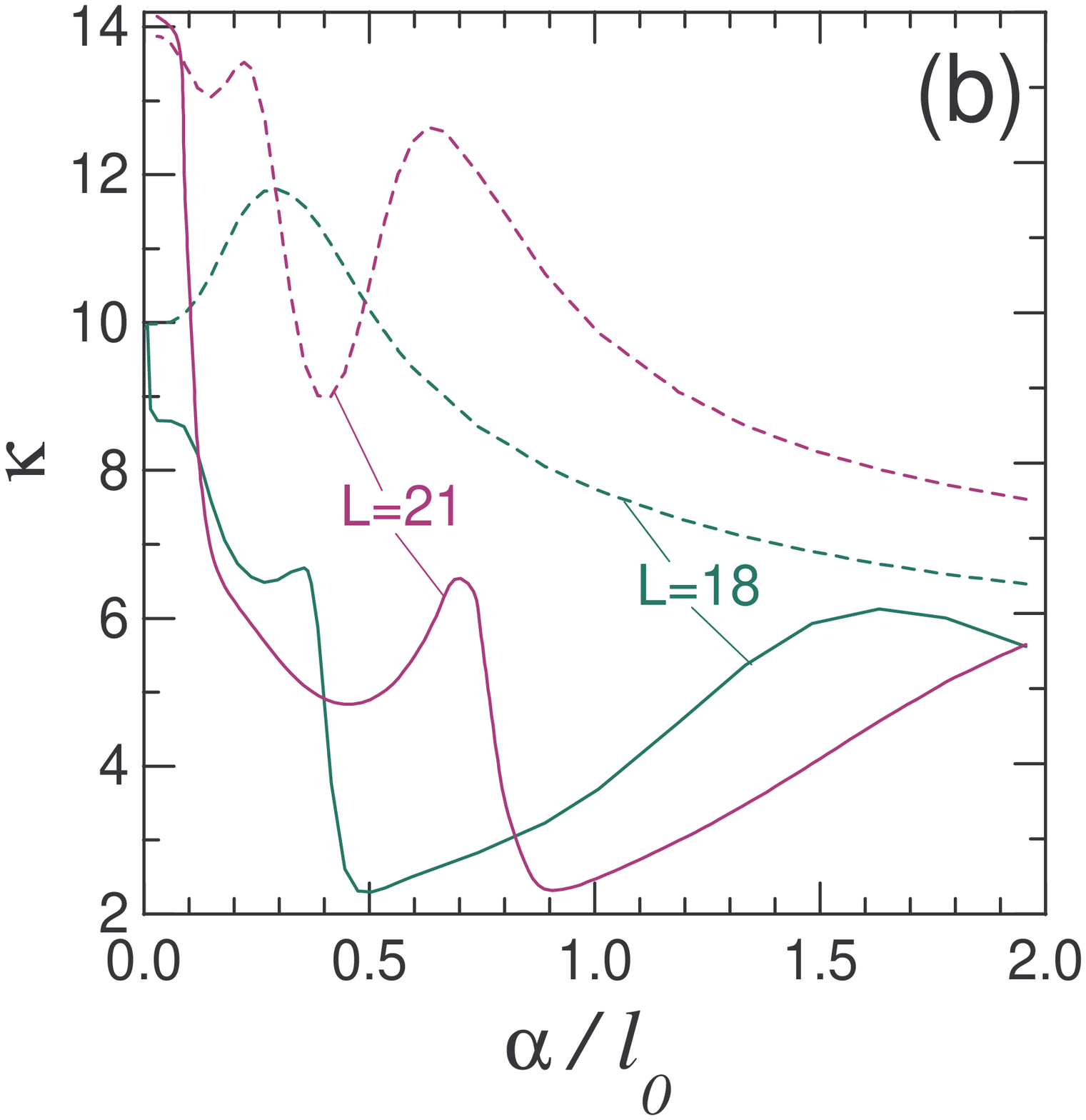}\hfill}
\caption{(color online) Coefficient $\kappa$ determining the
asymptotic dependence of the pair correlation function
 $W(z_a,z_b)\simeq|z_a-z_b|^\kappa$ for $z_a\rightarrow z_b$
 (see the text) for different total angular momentum states.
Solid curves were calculated as fits to the actual PCF values
calculated on an arc of length $0.6 l_0$ for $|z_a|=1.5l_0$, and
the dashed curves on a line segment of length $0.2l_0$ with one of
the ends fixed at the origin for $z_a=0$.
 }
\label{pcff}
\end{figure}

Another relevant quantity to be discussed is the pair correlation
function (PCF), defined as
\begin{equation}
W(z_a,z_b)=<\Psi|\sum_{i\neq j}\delta(z_i-z_a)\delta(z_j-z_b)|\Psi>.
\end{equation}
PCF calculated for the Laughlin function (\ref{laughlin}) gives
(up to a normalization constant)
\begin{equation}
\label{w2}
\begin{array}{lll}
W_L(z_a,z_b)=&(z_a-z_b)^{2m}\exp[-(|z_a|^2+|z_b|^2)/l_0^2]\\
&\\
&\times \int dz(z_a-z)^{2m}(z_b-z)^{2m}\exp(-|z|^2/l_0^2).
\end{array}
\end{equation}
Therefore, at small interelectron distances ($z_a\rightarrow z_b$)
the pair correlation function will asymptotically behave as
$W_L(z_a,z_b)\sim (z_a-z_b)^{2m}$. We consider the PCFs with one
particle fixed in $|z_a|=1.5l_0$ as well as at the origin $z_a$=0.
In the case of $|z_a|=1.5l_0$ we calculated the PCF for the other
electron at the same distance from the origin (i.e. $|z_b|=1.5l_0$ )
along an arc of $0.6l_0$ length away from $z_a$. Then, we fitted the
results to a function of the form
$f(|z_a-z_b|)=a|z_a-z_b|^{\kappa}$. For the other considered pinned
position ($z_a=0$) we repeated this procedure moving from the origin
along a straight line of length $0.2l_0$. The obtained results are
shown in Figs. \ref{pcff}(a) and (b). We found that the value of the
exponent depends on the fixed electron position ($z_a$), which is
not the case for the Laughlin wave function. For $L=9$ [blue curves
in Fig. \ref{pcff}(a)] the fitted $\kappa$ value approaches 6 -- the
Laughlin limit -- monotonically with decreasing $\alpha$. For
$L=12$, i.e. a non-Laughlin state, also the value of 6 is obtained
in the $\alpha=0$ limit. In this case three vortices become
localized at the electron position [see Fig. 1(b)] like for $L=9$.
We also notice that the $\kappa$ fitted for different fixed electron
positions [black solid and black dashed lines in Fig. \ref{pcff}(a)]
are both equal to 6 around $0.45l_0$, i.e. when the intermediate
giant vortex is formed [see Fig. 1(b)]. The intermediate giant
vortex is therefore associated with the appearance of a position
independent $\kappa$, which is characteristic of the Laughlin wave
function. Note, that the $\kappa$ value calculated for $z_a=0$
between the intermediate ($\alpha=0.45 l_0$) and the final
($\alpha=0$) giant vortices possesses a local minimum. This minimum
is related to those vortices which initially are moving away from
the electron for $\alpha$ below the occurrence of the intermediate
giant vortex (see the dashed lines in Fig. 1(b)]. However, for the
$\kappa$ exponent calculated with $|z_a|=1.5l_0$ [black dashed line
in Fig. \ref{pcff}(a)] a maximum is observed below $\alpha=0.45l_0$.
For $L=15$ the intermediate giant vortex is formed around
$\alpha=l_0$ [Fig. 1(c)]. The $\kappa$ values fitted for the two
pinned electron positions approach one another near $\alpha=l_0$
[see the solid and dashed red curves in Fig. \ref{pcff}(a)], but the
$\kappa$ value for $|z_a|=1.5l_0$ is larger than 6. As before, a
more direct correspondence between the vortex positions and the
fitted $\kappa$ value is obtained for the electron pinned at the
origin. The loop that the two vortices perform in the $(\alpha,y)$
plane when the intermediate giant vortex decays into single vortices
[see Fig. 1(c)], has no effect on the $\kappa$ value for
$|z_a|=1.5l_0$. On the other hand the loop is translated into a
minimum of $\kappa$ as calculated for $z_a=0$ [see the red solid
line in Fig. \ref{pcff}(a)]. Both $\kappa$ values tend to the value
of the Laughlin function, i.e. to 10, in the $\alpha=0$ limit. This
limit is achieved  by $\kappa$ values calculated for $L=18$ as well,
since also here 5 vortices are found at the electron position in the
contact potential limit [see Fig. 1(d)]. Again, for $L=18$ the
$\kappa$ value calculated for $z_a=0$ is more sensitive to the
actual vortex behavior. Local maximum (slightly above 6) is obtained
[dashed curve in Fig. \ref{pcff} (b)] when the first intermediate
giant vortex is formed [$\alpha\simeq 1.6l_0$, see Fig. 1(d)].
Another maximum is observed for the second intermediate giant vortex
($\alpha\simeq 0.37l_0$). The value is now considerably larger than
6, which can be explained by the presence of the other vortices
localized in close proximity of the pinned electron. The third
intermediate giant vortex near $0.1 l_0$ gives a plateau near
$\kappa=8.5$, which then shoots up to 10, when the final giant
vortex is formed. For $L=21$ we observe again a local maximum in
$\kappa$ calculated for $z_a=0$ at $\alpha=0.7l_0$ -- an
intermediate giant vortex position [see Fig. 1(e)]. For $L=21$ we
actually do not observe the final giant vortex in the
$\alpha\rightarrow 0$ limit, due to the problem of degeneracy of the
ground-state as explained in Section II. However, the presented
$\kappa$ values and the vortex positions for small $\alpha$ for
which the ground-state is still nondegenerate, clearly indicate the
giant vortex Laughlin asymptotic with seven vortices at the position
of the electron.

The close correspondence found between the PCF calculated for
$z_a=0$ and the vortex behavior is quite remarkable. For $L$ above
the value for the maximum density droplet, the charge density
develops a minimum at the center of the quantum dot and the depth
of the minimum increases with $L$. Moreover, by fixing the
position of one of the electrons at the origin one includes only
those Slater determinants in which the zero angular momentum
Fock-Darwin state appears. The angular momenta of the two
remaining orbitals must sum up to $l_a+l_b=L$ (let $l_a<l_b$).
From the asymptotic behavior of the single-electron orbitals at
the origin [$z^l$] one should expect that the obtained $\kappa$
value is related to the lowest of all $l_a$. Due to the applied
fitting procedure one actually obtains a $\kappa\geq l_a$. For
instance, the value of 6 obtained in the Laughlin limit for $L=9$
indicates that the Slater determinants corresponding to angular
momenta (0,1,8), and (0,2,7) do not contribute to the wave
function while (0,3,6) does. Expanding the Jastrow factor it is
straightforward to check that the Laughlin wave function contains
admixtures of the (0,3,6) and (0,4,5) basis functions, but not of
the (0,1,8) nor the (0,2,7) determinants.

\begin{figure*}[htbp]
\hbox{\epsfxsize=160mm
                \epsfbox[15 610 537 742] {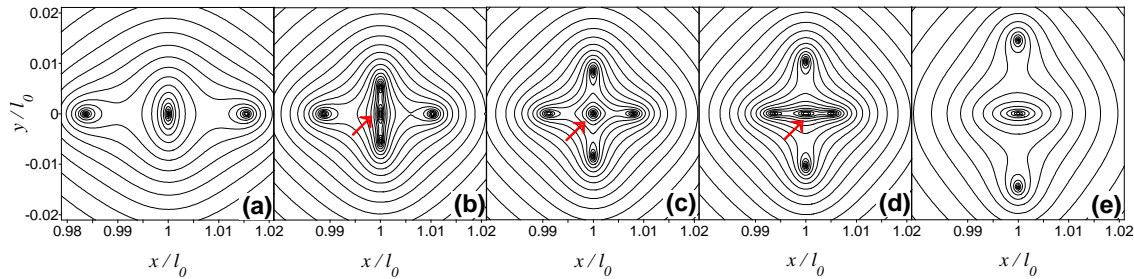}\hfill}
\caption{(color online) Contour plots of the logarithm of the
absolute value of the $L=12$ reduced wave function when the two
electrons are pinned at $(\pm l_0,0)$ calculated with a basis of
5158 Slater determinants including higher Landau levels. Plots (a-h)
correspond to the screening lengths $\alpha/l_0=0.421, 0.417, 0.416,
0.415$, and $0.412$, respectively. Red arrows point to the
antivortex positions. \label{av}
 }
\label{pcf}
\end{figure*}

\begin{figure}[htbp]
\epsfxsize=60mm
                \epsfbox[9 77 585 708] {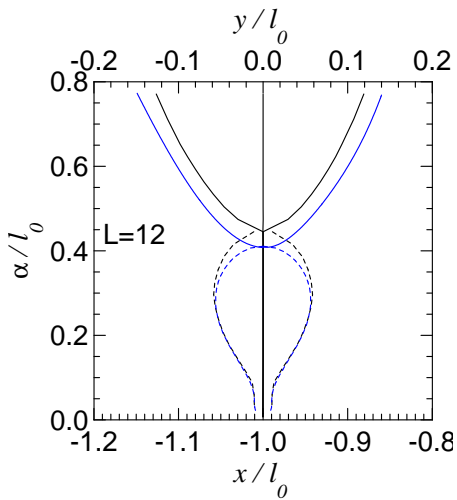}
\caption{(color online) Zoom of Fig. 1(b). Solid curves show the
$x$-position coordinates of the vortices localized at the $y=0$ axis
and the dashed lines are the $y$-position coordinates of vortices
localized at $x=\pm l_0$ line. Black curves correspond to the LLL
approximation and the exact results are plotted in blue.}
\end{figure}

\begin{table}
\begin{tabular}{ccccccc}
$E_{ni}$[meV] & $K$ & $E$[meV] & $x_l/l_0$ & $x_r/l_0$ & $\alpha^*/l_0$ & $c/l_0$ \\
  15      &  12   &  15.35272   &  -1.206    &  -0.813    &  0.440& - \\
  17      &  61   &  15.33831   &  -1.246    &  -0.779   &  0.450&  0.0042 \\
  19     &  173  &  15.33754   &  -1.243    &  -0.779   & 0.421 &  0.0073 \\
  21     &  392  &  15.33732   &  -1.238    &  -0.783   & 0.415 &  0.011 \\
  23     &  761  &  15.33722   &  -1.231    &  -0.786  &  0.411 &  0.014  \\
  25     &  1346 &  15.33719   &  -1.226   &  -0.789  & 0.409 & 0.016  \\
  27     & 2213  &  15.33717   &  -1.222    & -0.792   & 0.410 & 0.017  \\
  29     & 3453 & 15.33717  & -1.220 & -0.794&        0.413& 0.017 \\
  31     & 5158 & 15.33716 & -1.223 & -0.795 & 0.416 & 0.016 \\
\end{tabular}
\caption{Convergence of the results for $L=12$ beyond the LLL
approximation. $E_{ni}$ is the maximum energy of the noninteracting
Slater determinants used for the construction of the basis set
($B=0$). Second column lists the number of basis elements ($K$). $E$
is the energy estimate for $\alpha=1.48l_0$. $x_l$ and $x_r$ are the
positions of the bound vortices to the left and right of the
electron fixed at the point $(-l_0, 0)$ as in Fig. 1(b) for
$\alpha=1.48 l_0$. $\alpha^*$ is the screening length for which the
distance between the vortices aligned in the horizontal and vertical
directions is the same [see Fig. 8 (c)]. This distance ($c$) is
listed in the last column. The first row of the Table are the
results for the LLL approximation. Value for $\alpha^*$ in the first
row corresponds to the giant vortex.}
\end{table}

\subsection{Beyond the lowest Landau level approximation}
In order to verify the calculated vortex structure in the
neighborhood of the fixed electron we have performed exact
calculations with a basis including higher Landau levels for $L=12$.
The basis was constructed in the following way. From all the Slater
determinants built of the non-interacting Fock-Darwin states we
picked up only those for which the energy at $B=0$ (see the
discussion of the wave function scalability with the magnetic field
given in Section II) does not exceed a fixed energy value $E_{ni}$.
The number of basis elements $K$ as function of $E_{ni}$ is listed
in Table I together with the energy estimates obtained for an
interacting system at $\alpha=1.48 l_0$. The first row of the Table
corresponds to the LLL approximation. We obtain convergence of the
energy estimate up to six significant digits. Fourth and fifth
columns of the Table give the position of the vortices attached to
the electron localized at point $(-l_0,0)$ for the second electron
pinned at $(l_0,0)$ like in Fig. 1. The convergence of the position
of the vortices is slower than the energy. Beyond the LLL
approximation for $\alpha=1.48 l_0$ and for $\alpha$ up to the
Coulomb limit the distances between the electrons and the vortices
are slightly larger than in the LLL approximation. The positions of
vortices obtained with the most precise calculations are shown by
the blue curves in Fig. 1(b).

Beyond the LLL approximation the wave function is nonanalytical and
the exact number of nodes in the whole complex plane is not known a
priori. However, we have found that within the range plotted in Fig.
1(b) the extra nodes in the exact calculations appear only within
the region where the LLL predicts the formation of the intermediate
giant vortex. Fig. \ref{av} shows the contour plots of the logarithm
of the absolute value of the reduced wave function when the two
electrons are pinned at $(\pm l_0,0)$, for the range of $\alpha$
when the bound vortices flip their positions from the $x$ axis to
the $x=\pm l_0$ lines. Instead of the formation of the intermediate
giant vortex, a state consisting of single separate vortices is
formed. When the vortices approach the electron along the $x$ axis
[Figs. \ref{av}(a)], the node of the wave function associated with
the electron elongates in the perpendicular direction and finally
splits into an antivortex localized at the electron position and two
vortices localized at the $x=\pm l_0$ lines [Figs. \ref{av}(b)]
placed symmetrically with respect to the electron position. For a
certain screening length $\alpha=\alpha^*$ the distances between the
vortices localized at the $x$ axis and those localized at $x=\pm
l_0$ lines are equal [Fig. \ref{av}(c)].
 With decreasing $\alpha$ the
vortices localized at the $x$-axis approach [Fig. \ref{av}(d)] the
pinned electron and annihilate with the antivortex localized
therein. Eventually, we are left with a single vortex at the
electron position and two vortices localized on the $x=\pm l_0$ line
[Fig. \ref{av}(e)], like in the LLL for $\alpha$ values such that we
are between the intermediate and the final giant vortices. This
mechanism of the flip of the orientation of the vortices is found
for all the wave functions calculated beyond the LLL with basis
adopted according to the strategy explained above. Values of
$\alpha^*$ are listed in Table I. The corresponding distances
between the pairs of vortices ($c$) are given in the last column of
the Table. Distance $c$ initially increases with the size of the
variational basis and finally saturates near $0.016 l_0$.

Fig. 9 presents a zoom of Fig. 1(b) for the range of $\alpha$
corresponding to the intermediate and final giant vortices. The blue
curves are for the exact calculations. After the flip of the vortex
orientation the results of the LLL and the exact calculations are
nearly equal. Contribution of the higher LL becomes negligible when
the electron-electron interaction is switched off.

\subsection{Five electrons}
The mechanism presented above for the flip of the vortex orientation
is reproduced for higher number of electrons. To illustrate this we
focused on the five-electron system at $L=35$, i.e., a non-Laughlin
state corresponding to a ground-state of the magic angular momentum
sequence below the filling factor $\nu<1/3$. This state is the
counterpart of the $L=12$ state for three electrons discussed in the
context of Fig. 1(b). Calculations were performed in the LLL
approximation. The plots of the logarithm of the absolute value of
the reduced wave function are given in Fig. 10 for four electrons
fixed at the corners of a square $(\pm l_0,\pm l_0)$. Fig. 10(a)
shows the case of the Coulomb potential, and Figs. 10(b-d) the case
of the screened Coulomb interaction for $\alpha=0.0889l_0$, $0.0643
l_0$ and $0.0222l_0$. In Figs. 10(b-d) we present the vortices near
the electron localized at $(l_0,l_0)$. The vortices attached to the
fixed electrons approach them along the diagonals of the square and
form a giant vortex for $\alpha=0.0643 l_0$ [see Fig. 10(c)]. For
smaller values of $\alpha$ the line along which the attached
vortices are aligned is rotated over $90^\circ$ as compared to Fig.
10(b) and is now perpendicular to the corresponding diagonal of the
square [see Fig. 10(d)].

\begin{figure}[htbp]
\epsfxsize=90mm
                \epsfbox[37 330 520 812] {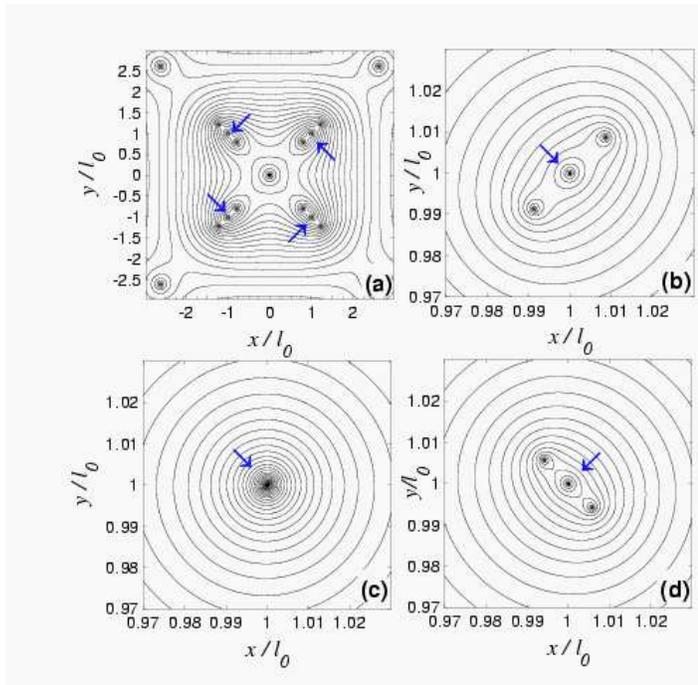}
\caption{(color online) Contour plots of the logarithm of the
absolute value of the reduced wave function calculated for five
electrons at angular momentum $L=35$ in the LLL approximation. Plot
(a) corresponds to the Coulomb interaction potential and plots
(b,c,d) to the screening lengths $\alpha=0.0889 l_0$ $\alpha=0.0643
l_0$ and $\alpha=0.0222 l_0$. Positions of four electrons are pinned
at the corners of the square $(\pm l_0,\pm l_0)$. Electron positions
are marked by blue arrows. }
\end{figure}

\section{Summary and Conclusions}

We have investigated the dependence of the vortex structure of a
three-electron quantum dot on the range of the inter-electron
potential. The Yukawa interaction potential can be changed
continuously from the Coulomb to the contact potential Laughlin
limit. The evolution towards the Laughlin liquid appears through the
formation of intermediate three-fold giant vortices at which the
vortices flip their orientation with respect to the electron to
which they are bound. In our discussion we relied on the reduced
wave function where $2$ electrons are pinned and found that the
screening lengths for which the giant vortices are formed do not
depend on the choice of the positions of the pinned electrons.
Hence, for $N=3$ the giant vortices can only be created by
manipulating the screening length and not the positions of the fixed
electron. For $N>3$ electrons the exact vortex structure in the
reduced wave function depends on the shape of the polygon formed by
the $N-1$ fixed electrons. But the binding of the vortices to the
fixed electrons for large L are independent of the exact location of
the electron. It is the angular position of the bound vortices which
is altered when we move the other fixed electrons. Nevertheless for
$N>3$, we find that the evolution to the Laughlin limit is also
non-monotonic and is accompanied with flips of the vortex
orientation and the formation of the intermediate composite fermion
states.

We found that the LLL approximation predicts the vortex positions
quite accurately in the whole range of the screening length except
for $\alpha$ values where the vortices approach closely the fixed
electrons. For a certain value of the screening length we observe a
flip of the vortex orientation. In general we found that this flip
can be realized in four different ways: symmetry breaking,
discontinuously, through a giant vortex or by the formation of
antivortices. In the LLL approximation giant vortices (similar to
the ones assumed in the Laughlin state) are observed at the
orientation flip, even though vortices are expected to exhibit a
repulsive behavior at close distances.\cite{wstep} In the LLL
approximation an antivortex can not appear because the number of
zeroes of the reduced wave function is fixed. When higher Landau
levels are included extra vortices and an antivortex appear and
annihilate preventing the formation of the giant vortex.

The presented study of the pair-correlation function shows that the
precise positions of the vortices with respect to the electrons are
important for the physics of electron-electron correlations. The
number of bound vortices in the close neighborhood of the electron
is translated into an asymptotic power-law form for the pair
correlation function around the pinned electron position. For the
giant vortices, i.e. for the intermediate composite vortex states,
the electron-electron correlations acquire properties similar to the
ones described by the Laughlin wave function.

 {\bf Acknowledgments}
This work was supported by the Flemish Science Foundation (FWO-Vl)
and the Belgian Science Policy. T.S was supported by the Marie
Curie training project HPMT-CT-2001-00394 and B.S by the EC Marie
Curie IEF project MEIF-CT-2004-500157.


\begin{thebibliography}{00}
\bibitem{uwaga} For $B\neq0$ the oscillator length $l_0$ should be
replaced by $l=\sqrt{\hbar/m^*\omega_e}$, with the effective
frequency $\omega_e^2=\omega^2+\omega_c^2/4$.
\end{thebibliography}
\end{document}